\documentclass[doublecolumn,11pt]{IEEEtran}
\usepackage{cite}
\usepackage{amsmath,amssymb,amsfonts}
\usepackage{graphicx,color}
\usepackage{textcomp}
\usepackage{hyperref}
\usepackage{enumitem}
\usepackage{multirow}
\usepackage{titlesec}
\usepackage{xcolor}
\usepackage{subcaption}
\usepackage{svg}
\usepackage{placeins}
\usepackage[linesnumbered,ruled,vlined]{algorithm2e}

\def\BibTeX{{\rm B\kern-.05em{\sc i\kern-.025em b}\kern-.08em
    T\kern-.1667em\lower.7ex\hbox{E}\kern-.125emX}}
\begin{document}

\title{
\textbf{From Electrochemical Energy Storage to Next-Generation Intelligent Battery Technologies for Electric Vehicles: A Survey}
}

\author{
\begin{center}
{Abderaouf Bahi\textsuperscript{\textbf{1*}}},
{Amel Ourici\textsuperscript{\textbf{2}}},
{Chaima Lagraa\textsuperscript{\textbf{3}}},
{Siham Lameche\textsuperscript{\textbf{3}}},
{Soundess Halimi\textsuperscript{\textbf{3}}}\\
{Inoussa Mouiche\textsuperscript{\textbf{4}}},
{Ylias Sabri\textsuperscript{\textbf{5,6}}},
{Waseem Haider\textsuperscript{\textbf{7}}} and
{Mohamed Trari\textsuperscript{\textbf{8}}} 
\\[1.2em]

\textsuperscript{\textbf{1}}\small
Computer Science and Applied Mathematics Laboratory (LIMA)\\ Faculty of Science and Technology, Chadli Bendjedid University, P.O. Box 73, El Tarf 36000, Algeria \\[0.6em]
\textsuperscript{\textbf{2}}\small
Mathematical Modeling and Numerical Simulation Laboratory (LAM2SIN)\\ Faculty of Technology, Badji Mokhtar University, P.O. Box 12, Annaba 23000, Algeria. \\[0.6em]
\textsuperscript{\textbf{3}}\small
Laboratory of Electrochemistry, Corrosion, Metallurgy and Mineral Chemistry (LECMCM)\\ Faculty of Chemistry, USTHB, Bab-Ezzouar 16111, Algiers, Algeria \\[0.6em]
\textsuperscript{\textbf{4}}\small
School of Computer Science, University of Windsor, ON, Canada \\[0.6em]
\textsuperscript{\textbf{5}}\small
School of Engineering, RMIT University, 124 La Trobe Street, 3001 Melbourne, Victoria, Australia \\[0.6em]
\textsuperscript{\textbf{6}}\small
Centre for Advanced Materials and Industrial Chemistry (CAMIC) \\ School of Science, RMIT University, Melbourne, Victoria 3001, Australia \\[0.6em]
\textsuperscript{\textbf{7}}\small
School of Engineering and Technology, Central Michigan University Mount Pleasant, MI 48859 USA
\\[1em]
\textsuperscript{\textbf{8}}\small
Laboratory of Storage and Valorization of Renewable Energies (LSVER)\\ Faculty of Chemistry, USTHB, Bab-Ezzouar 16111, Algiers, Algeria
\\[1em]

\textit{*Corresponding author:} Abderaouf Bahi (\textbf{a.bahi@univ-eltarf.dz})
\end{center}
}

\maketitle

\begin{abstract}
 This study provides a comprehensive overview of recent advances in electrochemical energy storage, including Na$^{+}$-ion, metal-ion, and metal-air batteries, alongside innovations in electrode engineering, electrolytes, and solid-electrolyte interphase control. It also explores the integration of machine learning, digital twins, large language models and predictive analytics to enable intelligent battery management systems, enhancing performance, safety, and operational longevity. Key challenges, research gaps, and future prospects are addressed, highlighting opportunities presented by hybrid chemistry, scalable manufacturing, sustainability, and AI-driven optimization. This survey aims to provide researchers, engineers, and industry profesionnals with a comprehensive understanding  of next-generation battery technologies for the evolving electric vehicles sector.
\end{abstract}


\begin{table}[h!]
\centering
\caption*{List of Abbreviations}
\begin{tabular}{ll}
\hline
\textbf{Abbreviation} & \textbf{Meaning} \\
\hline
AIC          & Akaike Information Criterion \\
AI           & Artificial Intelligence \\
ARIMA        & AutoRegressive Integrated Moving Average \\
BMS          & Battery Management System \\
BTMS         & Battery Thermal Management System \\
C-rate       & Charge/Discharge Rate \\
CNN          & Convolutional Neural Network \\
DL           & Deep Learning \\
DT           & Digital Twin \\
EKF          & Extended Kalman Filter \\
EV           & Electric Vehicle \\
EVRP         & Electric Vehicle Routing Problem \\
GRU          & Gated Recurrent Unit \\
GPT          & Generative Pre-trained Transformer \\
HIF-PF       & High-Importance-Factor Particle Filter \\
ICV          & Intelligent Connected Vehicle \\
IoT          & Internet of Things \\
LCA          & Life Cycle Assessment \\
LFP          & Lithium Iron Phosphate \\
LIB          & Lithium-Ion Battery \\
LLM          & Large Language Model \\
LNMO         & Lithium Nickel Manganese Oxide \\
MAE          & Mean Absolute Error \\
MAPE         & Mean Absolute Percentage Error \\
MAB          & Metal-Air Battery \\
ML           & Machine Learning \\
MLLM         & Multimodal Large Language Model \\
MLP          & Multi-Layer Perceptron \\
MSIformer    & Multi-Time Scale Interval Transformer \\
NCM          & Nickel-Cobalt-Manganese  \\
NCM111       & Nickel-Cobalt-Manganese (1:1:1 composition) \\
NMC          & Nickel-Manganese-Cobalt \\
PSO          & Particle Swarm Optimization \\
Q-learning   & Reinforcement Learning Algorithm \\
RAG          & Retrieval-Augmented Generation \\
RL           & Reinforcement Learning \\
RMSE         & Root Mean Square Error \\
RUL          & Remaining Useful Life \\
SIB          & Sodium-Ion Battery \\
SOC          & State of Charge \\
SOE          & State of Energy \\
SOH          & State of Health \\
SOP          & State of Power \\
\hline
\end{tabular}
\end{table}


\def\BibTeX{{\rm B\kern-.05em{\sc i\kern-.025em b}\kern-.08em
    T\kern-.1667em\lower.7ex\hbox{E}\kern-.125emX}}

\begin{figure}[!htbp]
\centering
\includegraphics[width=0.9\linewidth, height=1.0\textheight, keepaspectratio]{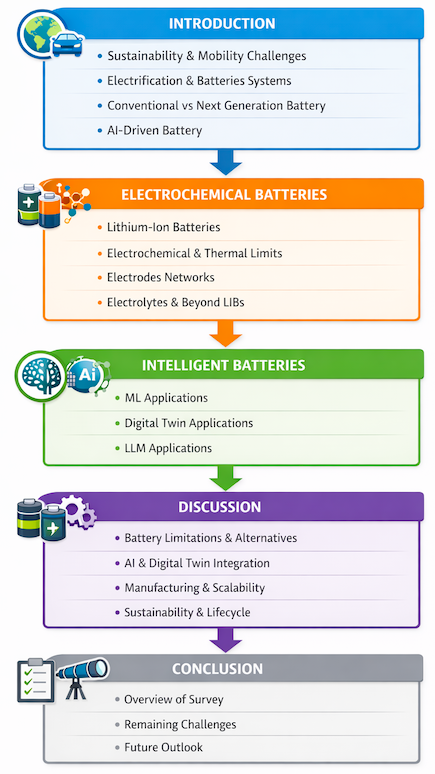}
\caption{Survey structure}
\end{figure}

\section{\textbf{Introduction}}
The global transition toward sustainable and low-carbon transportation has become one of the most critical challenges of the 21st century. Transportation alone accounts for a significant portion of global greenhouse gas emissions, air pollution, and fossil fuel consumption, prompting governments, industries, and research communities worldwide to accelerate the electrification of mobility systems \cite{AboKhalil2022,Rajper2020,Chen2022,Koech2020,Yang2022EV,Alanazi2023}. EVs have emerged as a cornerstone of this transformation, supported by rapid advancements in power electronics, charging infrastructure, and, most importantly, electrochemical energy storage technologies \cite{Lipu2022,How2019,Hannan2018}. As a result, the global EV market has experienced unprecedented growth over the past decade, with projections indicating a continued exponential increase in adoption across passenger vehicles, public transportation, and commercial fleets \cite{Sriram2022,Altenburg2014,Hossain2022,Deng2020,Mastoi2022,Sanguesa2021,Pipitone2021,JannesarNiri2024,Irle2023,Lu2024,IEA2023,King2023,GoldmanSachs2023,Lindwall2022}.

At the heart of this electrification revolution lies the battery system, which fundamentally determines vehicle range, safety, cost, reliability, and overall user acceptance \cite{Tomaszewska2019,Zhao2022}. Among existing technologies, Li-ion batteries have dominated the EV market due to their relatively high energy density, long cycle life, and technological maturity \cite{Wang2015,Czerwinski2021,Grant2022}. However, despite their widespread deployment, conventional Li-ion batteries are increasingly facing critical limitations that challenge their suitability for the next phase of large-scale EV adoption. These limitations include safety concerns related to thermal runaway \cite{Cheng2021}, high production and recycling costs \cite{Lipu2022}, dependence on scarce and geopolitically sensitive raw materials \cite{Yang2022EV}, aging and degradation issues \cite{Shah2022}, and limited adaptability to various driving conditions and user behaviors \cite{Hannan2018,Zubi2018,Zhang2018SOC}.

With the continued growth of EVs, these
challenges are intensifying, revealing a growing gap between the capabilities of conventional battery technologies and the demanding requirements of future electric mobility \cite{Zhao2022}. This gap has fueled an intensive global research effort aimed at developing next-generation battery technologies that go beyond incremental improvements \cite{Yang2022EV,Alanazi2023}. Emerging electrochemical systems such as solid-state batteries, lithium–sulfur, lithium–air, sodium-ion, and multivalent batteries promise higher energy densities, enhanced safety, lower environmental impact, and improved sustainability \cite{Mohammad2021}. In parallel, advances in materials science, nanotechnology, and electrochemistry are enabling novel electrode architectures and electrolyte formulations that redefine battery performance boundaries \cite{Tomaszewska2019,Grant2022}.

Beyond purely electrochemical innovations, a paradigm shift is occurring toward \emph{intelligent battery systems}. Modern EV batteries are no longer passive energy storage units but complex cyber physical systems tightly integrated with sensors, control units, and software layers. The rapid surge of AI, machine learning, and data-driven modeling have profoundly transformed BMS, enabling real time state estimation, predictive health monitoring, adaptive control, and fault diagnosis \cite{Lipu2022,Zhang2018SOC}. AI-driven BMS architectures leverage large-scale operational data to improve SOC, SOH, and RUL estimation, while optimizing safety, efficiency, and longevity under dynamic operating conditions \cite{Liu2022BMS}. Furthermore, hybrid energy storage systems combining batteries with supercapacitors or other storage technologies are gaining attention as a means to balance energy and power demands in EV applications \cite{Ali2023}.

Several surveys and review papers have investigated specific aspects of EV battery technologies, including electrochemical advancements, battery degradation mechanisms, safety analysis, and conventional BMS design \cite{Tomaszewska2019,Zhao2022,Shah2022}. Other studies have focused on AI applications in energy storage or intelligent control strategies for EVs \cite{Liu2022BMS}. However, existing surveys often address these topics in isolation, lacking a unified perspective that bridges the historical evolution of electrochemical energy storage with the emerging paradigm of intelligent, AI-enhanced battery systems \cite{Yang2022EV,Alanazi2023}. Moreover, a comprehensive synthesis that jointly examines materials innovation, next-generation battery chemistries, hybrid storage architectures, and intelligent management frameworks within a single coherent narrative remains limited \cite{Zhao2022,Wang2015}.

Motivated by this gap, this survey provides a comprehensive and forward-looking overview of battery technologies for electric vehicles, tracing their evolution from traditional electrochemical energy storage systems to next-generation intelligent battery technologies. To ensure methodological rigor and reproducibility, a systematic literature search was conducted across major scientific databases, including \textit{Web of Science}, \textit{Scopus}, \textit{IEEE Xplore}, \textit{ScienceDirect}, and \textit{Google Scholar}. The search strategy combined domain-specific keywords such as \textit{electric vehicle batteries}, \textit{advanced battery chemistries}, \textit{battery management systems}, \textit{AI-driven BMS}, and \textit{intelligent energy storage}, using Boolean operators to refine relevance.

This survey is based on more than 30 high-quality survey articles and over 150 peer-reviewed research papers published between 2023 and 2025, ensuring an up-to-date coverage of recent advances and emerging trends in the field \cite{AboKhalil2022,Rajper2020,Chen2022,Koech2020,Yang2022EV,Alanazi2023}. Inclusion criteria were defined to retain studies that (i) focus explicitly on EV battery technologies, (ii) present experimental, modeling, or system-level contributions, and (iii) are published in indexed journals or reputable international conferences. Studies lacking technical depth, industrial relevance, or methodological clarity were excluded.

Specifically, this survey first examines the development of conventional battery systems deployed in early and current EV platforms, followed by a systematic analysis of emerging battery chemistries and advanced materials aimed at overcoming limitations related to energy density, safety, cost, and sustainability \cite{Mohammad2021,Zhao2022}. In addition, strong emphasis is placed on intelligent battery management, encompassing AI-driven battery management systems, data-centric modeling techniques, predictive maintenance strategies, and hybrid energy storage architectures \cite{Shah2022,Liu2022BMS}. By integrating electrochemical, technological, and intelligent perspectives, this survey offers a holistic and structured understanding of the future trajectory of EV battery technologies \cite{Yang2022EV,Alanazi2023}.

The remainder of this paper is structured as follows. Section~2 reviews conventional battery technologies for electric vehicles, with a focus on classical electrochemical energy storage systems, their operating principles, advantages, and inherent limitations \cite{Tomaszewska2019,Wang2015,Grant2022}. Section~3 explores next-generation intelligent battery technologies, encompassing advanced battery chemistries, novel materials, AI-driven battery management systems, and hybrid energy storage architectures \cite{Shah2022,Liu2022BMS,Mohammad2021,Zhao2022}. Section~4 provides a critical discussion of open challenges, research gaps, and future directions, addressing issues related to scalability, safety, sustainability, standardization, and real-world deployment \cite{Yang2022EV,Alanazi2023}. Finally, Section~5 concludes with a summary of
key findings and the presentation of promising research avenues for the development of intelligent and sustainable battery systems that will shape the next generation of EVs\cite{AboKhalil2022,Rajper2020,Chen2022}.

\begin{table*}[!htbp]
\centering
\small
\caption{Comparison of Related Survey and Review Works on Electric Vehicle Batteries}
\label{tab:related_surveys}
\begin{tabular}{p{3.5cm} c p{6.5cm} p{5.5cm}}
\hline
\textbf{Authors} & \textbf{Year} & \textbf{Key Contributions} & \textbf{Limitations} \\
\hline

Zhang \textit{et al.} \cite{Zhang2025} & 2025 &
Comprehensive vision of smart batteries integrating advanced materials, sensing technologies, and artificial intelligence. &
Focuses on conceptual integration; lacks quantitative comparison of AI-driven BMS performance in EV-scale applications. \\

Madani \textit{et al.} \cite{Madani2025} & 2025 &
Detailed survey on AI and digital twin technologies for lithium-ion battery management, including state estimation. &
Limited discussion on industrial deployment challenges and real-world validation datasets. \\

Kurkin \textit{et al.} \cite{Kurkin2025} & 2025 &
Extensive review of battery management system circuitry configurations and algorithmic architectures for EVs. &
Primarily hardware and control oriented; minimal coverage of learning-based intelligence. \\

Ghazali \textit{et al.} \cite{Ghazali2025} & 2025 &
Systematic overview of advanced algorithms for EV battery management, emphasizing optimization and AI-based methods. &
Does not address battery chemistry evolution or thermal–electrochemical coupling. \\

Cavus \textit{et al.} \cite{Cavus2025} & 2025 &
Explores AI-driven predictive maintenance and intelligent battery management for next-generation EVs. &
Mainly conceptual; lacks detailed benchmarking across battery chemistries. \\

Huang \textit{et al.} \cite{Huang2025} & 2025 &
Comprehensive review of AI-based energy management strategies in hybrid and electric vehicles. &
Battery level degradation and electrochemical modeling are not deeply analyzed. \\

Tarout \textit{et al.} \cite{Tarout2025} & 2025 &
Bibliometric and systematic review on EV energy optimization techniques, highlighting recent research trends. &
Focuses on energy consumption rather than battery material and system-level intelligence. \\

Wadekar \textit{et al.} \cite{Wadekar2025} & 2025 &
Survey of smart charging and battery management technologies, including grid interaction and future EV ecosystems. &
Limited treatment of battery internal states and advanced sensing mechanisms. \\

Deng \textit{et al.} \cite{Deng2025} & 2025 &
Empirical analysis of socio technical factors influencing large scale EV adoption using real-world data. &
Does not address battery technology or management system design. \\

M.M. Hasan \textit{et al.} \cite{Hasan2025} & 2025 & Reviews advances and future directions of lithium-ion battery technologies, including materials, energy density improvements. & Focuses on technology trajectory; limited discussion on cost, commercialization challenges. \\

Lyu \textit{et al.} \cite{Lyu2024} & 2024 &
Comprehensive survey on intelligent BMS design, datasets, algorithmic frameworks, and emerging trends. &
Limited discussion on battery material evolution and solid-state technologies. \\

Bamdezh \textit{et al.} \cite{Bamdezh2024} & 2024 &
In-depth review of conventional and hybrid battery thermal management systems for EVs. &
Thermal focus only; electrical and AI-based management aspects are not covered. \\

Vikram \textit{et al.} \cite{Vikram2024} & 2024 &
Analysis of recent advancements and performance impacts of hybrid BTMS architectures. &
Does not integrate electrochemical degradation or intelligent control strategies. \\

Koech \textit{et al.} \cite{Koech2024} & 2024 &
Broad review of recent improvements in EV battery technologies across materials and systems. &
Lacks structured comparison of intelligent battery management approaches. \\

Tian \textit{et al.} \cite{Tian2024} & 2024 &
Critical analysis of inconsistency mechanisms and mitigation strategies in lithium-ion battery packs. &
Focuses on pack-level inconsistency without addressing AI-based diagnostics. \\

Celadon \textit{et al.} \cite{Celadon2024} & 2024 &
Integrated discussion of EV battery technology advancements, environmental impacts, and market trends. &
Limited technical depth on battery intelligence and control systems. \\

Iqbal \textit{et al.} \cite{Iqbal2023} & 2023 &
Survey of battery technologies and modeling methods for EV applications. &
Primarily modeling oriented; minimal focus on AI-enhanced management systems. \\

Abro \textit{et al.} \cite{Abro2023} & 2023 &
Comprehensive review of battery, propulsion, and vehicle network technologies for intelligent EVs. &
Battery chemistry evolution and degradation mechanisms are only briefly discussed. \\

Manoj \textit{et al.} \cite{Manoj2023} & 2023 &
Review of optimization and AI algorithms for effective battery management in EVs. &
Does not link algorithmic methods with battery material constraints. \\

\hline
\end{tabular}
\end{table*}

\section{\textbf{Electrochemical Batteries for Electric Vehicles}}
\subsection{\textbf{Lithium-Ion Batteries: Materials, Microstructure, and Electrochemical Limitations}}

\begin{figure*}[!htbp]
\centering
\includegraphics[width=\textwidth, height=0.7\textheight, keepaspectratio]{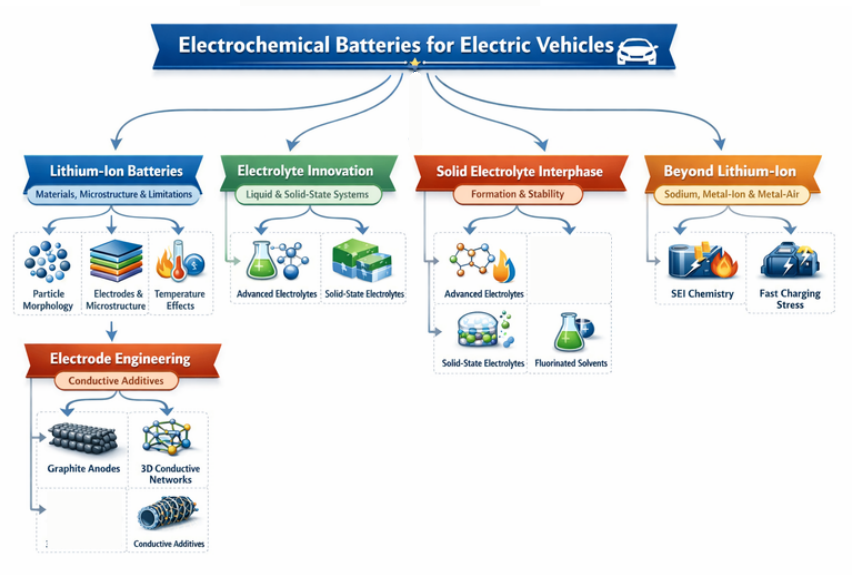}
\caption{Taxonomy of electrochemical batteries for electric vehicles}
\label{fig:taxonomy_batteries}
\end{figure*}

Lithium ion batteries remain the dominant energy storage technology for electric vehicles due to their high energy density, mature manufacturing processes, and reliable electrochemical performance. Nevertheless, their large scale deployment under real world driving and charging conditions reveals persistent limitations associated with materials degradation, electrochemical instability, and thermal sensitivity. Addressing these challenges requires a comprehensive understanding of aging mechanisms, electrode material behavior, electrolyte stability, and system level constraints.

At the cell aging level, performance degradation is strongly dependent on usage conditions and accumulated driving mileage. Through detailed electrochemical analysis of EV aged cells, Yap \textit{et al.} identified lithium inventory loss as the primary driver of early capacity fade, while active material loss progressively dominates degradation beyond high mileage operation \cite{Yap2025AgingEVBatteries}. Their results further indicate that cathode degradation constitutes the main limiting factor when the negative to positive electrode capacity ratio exceeds unity, emphasizing the critical role of cathode stability in long term EV battery performance.

From an electrode materials perspective, the pursuit of higher energy density has intensified research on advanced anode and cathode chemistries. Silicon based anodes, despite their exceptional theoretical capacity, suffer from severe volume expansion and unstable interfacial chemistry. Li \textit{et al.} systematically reviewed lithium storage and failure mechanisms in silicon based anodes, demonstrating that mechanical fracture, solid electrolyte interphase instability, and electrical isolation remain fundamental barriers to commercial adoption \cite{Li2025SiliconAnodes}. Their analysis highlights the necessity of structural optimization and composite electrode design to achieve cycling stability.

Cathode materials represent another critical bottleneck, particularly with respect to sustainability and resource efficiency. In this context, Wang \textit{et al.} proposed an effective regeneration strategy for spent NCM cathodes using in situ surface impurity conversion combined with protective coating formation \cite{Wang2025RecyclingNCM}. This approach not only restores electrochemical performance but also enables high value recycling of end of life battery materials, contributing to circular economy objectives. Complementary to recycling strategies, compositional tuning through cationic substitution has proven effective. Toghan \textit{et al.} demonstrated that Cs ion doping in high voltage LiNi$_{0.5}$Mn$_{1.5}$O$_4$ cathodes enhances ionic conductivity, structural stability, and capacity retention under both ambient and elevated temperature conditions \cite{Toghan2025CsDopedLNMO}.

Electrolyte behavior further constrains the electrochemical performance and safety of lithium ion batteries. Beyond conventional solvent salt optimization, lithiation based material strategies have emerged as a promising route to improve ionic transport and interfacial stability. Zardehi Tabriz \textit{et al.} reviewed the application of lithiated materials across electrolytes, separators, binders, and artificial interphases, demonstrating their potential to enhance lithium transference number, suppress dendritic growth, and improve cycling performance \cite{ZardehiTabriz2025LithiatedMaterials}. Despite these advances, electrolyte degradation and instability under high voltage operation remain unresolved challenges.

Fast charging operation introduces additional degradation pathways, most notably lithium plating on the anode surface. Shin and Lee developed an electrochemistry based diagnostic approach capable of identifying lithium plating induced degradation through characteristic current signatures during constant voltage charging \cite{Shin2025LiPlatingDetection}. Such methods are essential for distinguishing lithium plating from other aging mechanisms and for enabling safer fast charging strategies in electric vehicles.

Thermal effects strongly interact with electrochemical degradation processes, particularly at high charge and discharge rates. Advanced battery thermal management systems are therefore critical for ensuring operational safety and performance stability. Kang \textit{et al.} demonstrated that a coupled phase change material and liquid cooling system significantly improves temperature uniformity and suppresses peak temperatures in high rate battery packs \cite{Kang2025CoupledBTMS}. Their results highlight the necessity of integrated thermal-electrochemical co-design for EV battery systems.

Finally, material and electrochemical limitations must be considered at the energy system scale. Dey \textit{et al.} reviewed recent advancements in lithium ion battery materials for battery energy storage systems, emphasizing that material scarcity, safety constraints, and energy density limits motivate continued exploration of advanced materials and alternative chemistries \cite{Dey2025BESSMaterials}. This broader perspective reinforces the need for coordinated innovation across materials, cell design, and system integration to meet the future requirements of electric mobility.

\begin{table*}[!htbp]
\centering
\small
\caption{Key References on Lithium-Ion Batteries: Materials, Microstructure, and Electrochemical Limitations}
\label{tab:liion_subnode}
\begin{tabular}{p{3.5cm} c p{6.5cm} p{5.5cm}}
\hline
\textbf{Authors} & \textbf{Year} & \textbf{Key Contributions} & \textbf{Limitations} \\
\hline

Yap \textit{et al.} \cite{Yap2025AgingEVBatteries} & 2025 &
Experimental identification and quantification of real world aging mechanisms in EV lithium ion batteries, highlighting lithium inventory loss and cathode driven degradation as dominant factors. &
Focuses on post mortem and diagnostic analysis; mitigation strategies are not addressed. \\

Kang \textit{et al.} \cite{Kang2025CoupledBTMS} & 2025 &
Proposes a coupled phase change material and liquid cooling thermal management system that significantly improves temperature uniformity and safety under high discharge rates. &
System complexity and integration cost are not fully evaluated for large scale deployment. \\

Wang \textit{et al.} \cite{Wang2025RecyclingNCM} & 2025 &
Introduces an in situ regeneration and coating strategy for spent NCM cathodes, achieving high electrochemical performance and enabling sustainable battery recycling. &
Applicability to other cathode chemistries remains unexplored. \\

Li \textit{et al.} \cite{Li2025SiliconAnodes} & 2025 &
Comprehensive review of lithium storage mechanisms and failure modes in silicon based anodes, with emphasis on structural optimization strategies for cycling stability. &
Primarily review oriented; limited experimental validation of proposed designs. \\

Tabriz \textit{et al.} \cite{ZardehiTabriz2025LithiatedMaterials} & 2025 &
Systematic review of lithiated materials applied to electrolytes, separators, binders, and interphases to enhance ionic conductivity and battery performance. &
Manufacturing scalability and long term stability require further investigation. \\

Shin \textit{et al.} \cite{Shin2025LiPlatingDetection} & 2025 &
Development of an electrochemistry based diagnostic method to detect lithium plating induced degradation during fast charging operation. &
Method is validated under controlled conditions; real time onboard implementation is not addressed. \\

Toghan \textit{et al.} \cite{Toghan2025CsDopedLNMO} & 2025 &
Demonstrates performance enhancement of high voltage LiNi0.5Mn1.5O4 cathodes through Cs ion doping, improving ionic conductivity and cycling stability. &
Doping concentration optimization and long term thermal stability remain open issues. \\

Dey \textit{et al.} \cite{Dey2025BESSMaterials} & 2025 &
Review of recent advancements in lithium ion battery materials for battery energy storage systems, emphasizing energy density, safety, and sustainability challenges. &
Focuses on material trends; lacks detailed electrochemical degradation analysis. \\

Zhang \textit{et al.} \cite{Zhang2022Particuology} & 2022 &
Overview of lithium ion battery materials and particle level microstructure effects on electrochemical performance. &
Limited discussion on cell level aging and system integration. \\

Sarkar \textit{et al.} \cite{Sarkar2021Electrochimica} & 2021 &
Insights into electrode microstructure optimization and its influence on cycling performance and degradation behavior. &
Primarily laboratory scale; industrial applicability is not assessed. \\

Weiss \textit{et al.} \cite{Weiss2021AEM} & 2021 &
Analysis of fast charging related degradation mechanisms and material level strategies for performance improvement. &
Experimental validation under real world EV conditions is limited. \\

Alipour \textit{et al.} \cite{Alipour2020Batteries} & 2020 &
Comprehensive review of temperature dependent electrochemical behavior and safety challenges in lithium ion batteries. &
Does not address recent high energy density cathode materials. \\

\hline
\end{tabular}
\end{table*}

\subsection{\textbf{Electrode Engineering and Conductive Additives}}

Enhancing rate capability, fast charging tolerance, and long term cycling stability in lithium ion batteries relies critically on the rational engineering of electrode microstructures and the optimization of conductive additive systems. In composite electrodes, electrochemical performance is governed not only by the intrinsic properties of active materials but also by the efficiency of electronic and ionic transport pathways formed within the electrode matrix. As electrode thickness and mass loading increase to meet energy density requirements, transport limitations become increasingly severe, necessitating advanced conductive network designs.

Conventional carbon black additives provide localized electronic contacts but often fail to sustain efficient charge transport in thick or high mass loading electrodes. To overcome these limitations, recent studies have emphasized the importance of conductive additive morphology, defect density, and dimensionality. He \textit{et al.} demonstrated that reduced graphene oxide with an optimized defect concentration enables superior interfacial contact between active material particles and current collectors, facilitating simultaneous electron and lithium ion transport while preserving high rate capability \cite{He2025rGOAdditives}. Their results highlight that a balanced defect density is essential to avoid ion transport hindrance associated with highly ordered graphene structures.

Carbon nanotube based additives have also attracted significant attention due to their one dimensional geometry and excellent electrical conductivity. However, their tendency to form entangled agglomerates can limit effective network formation. Cho and Piao addressed this issue by introducing shortly cut carbon nanotubes, which form dense and well distributed conductive frameworks within silicon based anodes \cite{Cho2025ShortCNT}. This tailored architecture improves electronic percolation and accommodates volumetric expansion, resulting in enhanced rate performance and cycling stability.

Beyond purely electronic conduction, hybrid electronic ionic conductive additives have emerged as a promising strategy for fast charging applications. Chen \textit{et al.} proposed an electronic ionic hybrid additive composed of carbon black decorated with an ultrathin lithium phosphide layer formed in situ during battery operation \cite{Chen2025HybridAdditive}. This multifunctional additive significantly enhances lithium ion transport while maintaining electronic conductivity, enabling rapid capacity recovery even in thick electrodes with high areal loading.

In high nickel cathodes, where polarization and ohmic resistance become critical at elevated mass loadings, multidimensional conductive architectures have proven particularly effective. Saqib \textit{et al.} demonstrated that combining carbon black with carbon nanofibers creates synergistic conductive networks that minimize contact resistance and enhance lithium ion diffusion in nickel rich layered oxide cathodes \cite{Saqib2025MultidimensionalCarbon}. The long range connectivity provided by fibrous additives plays a key role in sustaining electrochemical performance under high current densities.

Conductive additive engineering is also essential in emerging electrode manufacturing technologies such as dry electrode processing. Hwang \textit{et al.} reported that spray dried single walled carbon nanotube coated cathode particles enable uniform conductive network formation with significantly reduced additive content, leading to excellent rate capability and long term cycling stability in dry processed electrodes \cite{Hwang2025DryElectrodeSWCNT}. This approach underscores the importance of additive dispersion and interfacial control in next generation battery fabrication routes.

Beyond synthetic carbon materials, biomass derived conductive additives have recently gained attention due to their structural tunability and sustainability. Ge \textit{et al.} developed nitrogen doped porous carbon derived from pine needles that forms three dimensional conductive frameworks while providing additional lithium storage sites \cite{Ge2025BiomassCarbon}. When combined with conventional carbon black, these biomass based additives significantly improve reversible capacity and rate performance.

Non carbon conductive additives have also been explored to improve electrode stability and high rate performance. Wang \textit{et al.} demonstrated that incorporating Ti$_4$O$_7$ as a conductive additive in NCM111 cathodes reduces polarization and enhances cycling stability, particularly under high current operation \cite{Wang2025Ti4O7Additive}. Such conductive oxides provide an alternative pathway for designing robust conductive networks in demanding electrochemical environments.

Overall, these studies demonstrate that conductive additives should be regarded as active design parameters rather than passive components. Future electrode engineering strategies will increasingly rely on multifunctional and hybrid conductive systems capable of simultaneously addressing electronic conduction, ionic transport, structural integrity, and manufacturability in high energy density lithium ion batteries.

\begin{table*}[!htbp]
\centering
\small
\caption{Key References on Electrode Engineering and Conductive Additives in Lithium-Ion Batteries}
\label{tab:electrode_subnode}
\begin{tabular}{p{3.5cm} c p{6.5cm} p{5.5cm}}
\hline
\textbf{Authors} & \textbf{Year} & \textbf{Key Contributions} & \textbf{Limitations} \\
\hline

He \textit{et al.} \cite{He2025rGOAdditives} & 2025 &
Demonstrated that reduced graphene oxide with optimized defect concentration enhances electron/ion transport and rate capability in LiFePO4 and LiCoO2 electrodes. &
Limited to specific cathode chemistries; full-cell performance not evaluated. \\

Cho \textit{et al.} \cite{Cho2025ShortCNT} & 2025 &
Introduced shortly cut carbon nanotubes to improve conductive networks in silicon anodes, enhancing rate performance and accommodating volumetric expansion. &
Focused on Si anodes; impact on full-cell systems not reported. \\

Chen \textit{et al.} \cite{Chen2025HybridAdditive} & 2025 &
Proposed electronic/ionic hybrid additive for thick electrodes, improving Li+ transport and fast charging capability while maintaining electronic conductivity. &
Requires complex additive synthesis; scalability and cost not addressed. \\

Saqib \textit{et al.} \cite{Saqib2025MultidimensionalCarbon} & 2025 &
Developed multidimensional carbon networks in Ni-rich cathodes to reduce contact resistance and enhance rate capability in high-mass loading electrodes. &
Limited investigation on cycling under varied temperature; primarily NCM cathodes studied. \\

Hwang \textit{et al.} \cite{Hwang2025DryElectrodeSWCNT} & 2025 &
Spray-dried SWCNT/NCM composites enabled uniform conductive networks with reduced additive content, improving rate performance and cycling stability. &
Study focused on dry electrode fabrication; applicability to wet process not discussed. \\

Lewis \textit{et al.} \cite{LewisAcidBase2025HighVoltageLIB} & 2025 &
Demonstrated multifunctional Lewis acid-base electrolyte additive and conductive network for high-voltage LNMO/graphite cells, enhancing cycling stability and rate capability. &
Publisher information pending; experimental details limited. \\

Ge \textit{et al.} \cite{Ge2025BiomassCarbon} & 2025 &
Used pine needle derived porous carbon as conductive additive to form 3D frameworks, enhancing reversible capacity, rate performance, and cycle life. &
Focus on electrode material optimization; full cell integration not fully demonstrated. \\

Wang \textit{et al.} \cite{Wang2025Ti4O7Additive} & 2025 &
Incorporated Ti$_4$O$_7$ as conductive additive in NCM111 cathodes, reducing polarization and improving high-rate cycling stability. &
Primarily evaluated in lab-scale NCM111; impact on other cathode chemistries not explored. \\

Ye \textit{et al.} \cite{Ye2023DRM} & 2023 &
Demonstrated that 3D and multidimensional conductive frameworks enhance electron percolation, Li-ion diffusion, rate performance, and structural robustness of graphite anodes. &
Focused on anode architectures; cathode and full-cell interactions not fully explored. \\

Jiao \textit{et al.} \cite{Jiao2023ACSAEM} & 2023 &
Systematic analysis of conductive additives’ impact on cathode electrochemical behavior, reducing polarization losses and enhancing high rate performance. &
Limited discussion on long-term cycling and full cell integration. \\

Kang \textit{et al.} \cite{Kang2022JPS} & 2022 &
Incorporated fiber shaped conductive additives to form long range conductive networks, improving electronic connectivity &
Primarily tested on specific anode/cathode systems; scalability not addressed. \\

\hline
\end{tabular}
\end{table*}

\subsection{\textbf{Electrolyte Innovation and the Transition Toward Solid-State Systems}}

The intrinsic safety risks associated with conventional flammable liquid electrolytes have motivated extensive research into advanced electrolyte formulations capable of supporting higher voltages, wider operating temperature ranges, and improved interfacial stability. Among the most promising approaches, fluorinated electrolytes have attracted significant attention due to their enhanced oxidative stability and thermal robustness. Zhang \textit{et al.} demonstrated that fluorinated electrolyte formulations enable stable battery operation beyond 5~V, thereby substantially expanding the electrochemical stability window of lithium-ion batteries \cite{Zhang2013EnergyEnvSci}. Subsequent studies confirmed that fluorinated solvents improve thermal tolerance, low temperature performance, and electrode electrolyte interfacial stability, particularly under high-voltage and high-rate conditions \cite{Lavi2020ACSAEM,Fan2019NatureEnergy}.

Beyond fluorination strategies, alternative solvent chemistries and solvation control have been explored to address the competing demands of safety, fast charging, and electrochemical performance. Nitrile-based electrolytes and tailored solvation structures have been shown to enhance thermal stability while enabling higher charging rates, mitigating degradation phenomena associated with aggressive operating conditions \cite{Kerner2016JPS,Kautz2023AEM}. From a broader perspective, Logan and Dahn articulated key electrolyte design principles for fast charging lithium-ion batteries, emphasizing the delicate balance between ionic conductivity, interfacial chemistry, and safety constraints \cite{Logan2020TrendsChem}. More recently, electrolyte optimization strategies have focused on stabilizing nickel rich cathode materials under high-voltage cycling, addressing critical degradation pathways that limit energy density and cycle life \cite{Zhao2023Batteries}.

Despite these advances, the fundamental limitations of liquid electrolytes including flammability, leakage risks, and restricted electrochemical stability have accelerated the transition toward solid-state battery architectures. Solid polymer and composite electrolytes offer inherent safety advantages, improved mechanical integrity, and compatibility with lithium metal anodes, making them attractive candidates for next-generation energy storage systems. Recent comprehensive reviews have systematically examined ion transport mechanisms, ionic conductivity bottlenecks, and material design strategies in solid-state lithium batteries, highlighting both their potential and remaining challenges \cite{Yang2022ESE,Cheng2021ESM,Hadad2025SolidStateRevolution,Jose2025SolidStateLIBs,Su2025SSBFabrication}.

Significant European efforts such as the HELENA project focus on the development of high energy density lithium-metal halide solid-state batteries for electric vehicles and aircrafts, advancing cell design with nickel-rich cathodes, lithium-metal anodes, and lithium-ion superionic halide solid electrolytes to maximize both energy and power density \cite{LopezAranguren2025HELENA}. These projects integrate innovations across material processing, cell fabrication, battery modelling, and recycling strategies, demonstrating the pathway from laboratory research to pre-industrial prototypes.

From a broader energy transition perspective, recent reviews have highlighted the critical role of solid-state electrolytes in enabling safer, high-performance batteries and in accelerating the adoption of electric mobility technologies. Advanced polymeric and composite electrolytes offer prospects for overcoming ionic transport limitations while maintaining mechanical stability, which is essential for both automotive and aerospace applications \cite{Malik2025EVTechnology}. Collectively, these studies underscore the ongoing shift toward solid-state battery architectures as a strategic direction for sustainable, high-performance energy storage solutions.
.

\begin{table*}[!htbp]
\centering
\small
\caption{Key References on Electrolyte Innovation and Transition Toward Solid State Systems}
\label{tab:electrolyte_subnode}
\begin{tabular}{p{3.5cm} c p{6.5cm} p{5.5cm}}
\hline
\textbf{Authors} & \textbf{Year} & \textbf{Key Contributions} & \textbf{Limitations} \\
\hline

Aranguren \textit{et al.} \cite{LopezAranguren2025HELENA} & 2025 &
Presented the HELENA project advancing high energy density Li-metal halide solid-state batteries for EVs and aircrafts, covering material processing, cell fabrication, battery modelling, and recycling strategies. &
Focus on pre-industrial prototypes; long-term operational performance not fully demonstrated. \\

Hadad \textit{et al.} \cite{Hadad2025SolidStateRevolution} & 2025 &
Reviewed solid polymer electrolytes, emphasizing ion transport mechanisms, ionic conductivity limitations, and interface engineering for safer lithium-ion batteries. &
Predominantly theoretical and lab-scale studies; practical integration challenges remain. \\

Jose \textit{et al.} \cite{Jose2025SolidStateLIBs} & 2025 &
Analyzed advances and challenges in solid-state lithium batteries, including materials for electrolyte, anode, cathode, and interface optimization strategies. &
Review; limited experimental validation for large-scale systems. \\

Su \textit{et al.} \cite{Su2025SSBFabrication} & 2025 &
Compared sulfide and oxide solid electrolytes, developed advanced fabrication techniques for hybrid solid-state batteries, and analyzed interfacial resistance contributions. &
Focused on specific fabrication techniques; scalability and mass production constraints need further study. \\

Malik \textit{et al.} \cite{Malik2025EVTechnology} & 2025 &
Provided comprehensive review of EV technology, emphasizing solid-state batteries, battery energy density improvements, fast-charging infrastructure, and sustainable mobility. &
Broad scope; detailed technical performance analysis of specific battery chemistries is limited. \\

Kautz \textit{et al.} \cite{Kautz2023AEM} & 2023 &
Investigated tailored solvation structures to mitigate degradation under aggressive operating conditions. &
Focused on electrolyte chemistry; full cell system interactions not fully addressed. \\

Zhao \textit{et al.} \cite{Zhao2023Batteries} & 2023 &
Proposed electrolyte optimization strategies to stabilize nickel rich cathodes under high voltage cycling. &
Mainly focused on cathode stabilization; broader battery systems not included. \\

Lin \textit{et al.} \cite{Lin2023MATEC} & 2023 &
Surveyed solid state sodium batteries as cost effective and safe alternatives for large-scale energy storage. &
Focuses on Na based systems; comparison with Li-based solid state electrolytes limited. \\

Yang \textit{et al.} \cite{Yang2022ESE} & 2022 &
Comprehensive review on ion transport, ionic conductivity bottlenecks, and material design in solid-state Li batteries. &
Emphasizes solid-state materials; practical implementation and cost analysis limited. \\

Cheng \textit{et al.} \cite{Cheng2021ESM} & 2021 &
Reviewed solid polymer and composite electrolytes for enhanced safety, mechanical integrity, and Li-metal compatibility. &
Mostly lab-scale studies; large-scale applicability not deeply addressed. \\

Lavi \textit{et al.} \cite{Lavi2020ACSAEM} & 2020 &
Confirmed that fluorinated solvents enhance electrode electrolyte interfacial stability under high-voltage and high rate conditions. &
Mainly cathode-focused; anode and full-cell performance less explored. \\

Logan \textit{et al.} \cite{Logan2020TrendsChem} & 2020 &
Outlined key design principles for fast-charging Li-ion batteries, balancing ionic conductivity, interfacial chemistry, and safety. &
Conceptual review; lacks quantitative benchmarking. \\

\hline
\end{tabular}
\end{table*}

\subsection{\textbf{Solid Electrolyte Interphase: Formation, Stability, and Performance Implications}}

The solid electrolyte interphase (SEI) constitutes a critical component governing the performance, safety, and lifetime of lithium-ion batteries. Formed primarily during initial formation cycles, the SEI acts as a passivating layer that enables lithium-ion transport while suppressing continuous electrolyte decomposition at the electrode surface. Despite decades of research, the physicochemical mechanisms underlying SEI formation, evolution, and degradation remain only partially understood, particularly under the demanding operating conditions of electric vehicles.

Early foundational studies established the conceptual framework of SEI chemistry, highlighting its dynamic and heterogeneous nature. The composition and stability of SEI are highly sensitive to electrolyte formulation, electrode surface chemistry, and cycling protocols. Subsequent reviews provided detailed analyses of SEI formation pathways, chemical constituents, and mitigation strategies, emphasizing the strong coupling between electrochemical reactions and interfacial phenomena \cite{An2016Carbon,Lin2020NanoEnergy}. Manthiram \textit{et al.} demonstrated that SEI instability becomes increasingly pronounced in high energy density batteries, where elevated voltages and aggressive charging accelerate interfacial degradation \cite{Manthiram2019AEM}.

Recent surveys have advanced the understanding of SEI evolution in graphite-based lithium-ion batteries. Adenusi \textit{et al.} reviewed SEI engineering strategies, including electrolyte additives and surface modifications, to stabilize the electrode-electrolyte interface \cite{Adenusi2023AEM}. Beheshti \textit{et al.} examined graphite electrolyte interphase formation, highlighting the role of additives in long-term electrochemical performance \cite{Beheshti2022iScience}. Complementary experimental and modeling studies identified distinct SEI growth regimes during cycling, which influence capacity fade and impedance rise \cite{Kolzenberg2020NCB}.

Mechanical and operational stresses further complicate SEI stability. Nie \textit{et al.} reported that SEI degradation significantly contributes to performance loss under fast-charging conditions, promoting lithium plating and interfacial instability \cite{Nie2018Angew}. Li \textit{et al.} highlighted the interplay between SEI mechanical properties and battery lifetime, showing that stress accumulation and fracture accelerate interfacial degradation \cite{Li2020CRPS,Li2024SEIMechanical}. Vibration-induced effects, as explored by Ryu \textit{et al.}, can selectively remove fragile organic components, yielding thinner and more inorganic-rich SEI layers that enhance Li$^+$ transport and battery performance \cite{Ryu2025VibrationSEI}.

Artificial SEI  strategies have emerged as effective approaches to enhance interphase stability. Zhang \textit{et al.} summarized recent advances in artificial SEI fabrication, emphasizing tailored compositions and microstructures that optimize ionic conductivity and structural robustness for high-energy batteries \cite{Zhang2025ArtSEI}. Similarly, Liang \textit{et al.} demonstrated that highly ductile dual-halide SEI layers on lithium metal anodes improve ionic transport, suppress dendrite growth, and maintain structural stability during volume changes \cite{Liang2025DualHalideSEI}. Hao \textit{et al.} provided a comprehensive overview of SEI structural evolution at electrode-electrolyte interfaces, elucidating mechanisms governing composition, growth, and degradation \cite{Hao2025SEIStructure}.

As battery technologies advance toward lithium metal anodes and solid-state architectures, SEI-related challenges persist in more severe forms. Solid-state batteries promise higher energy density and improved safety but introduce interfacial challenges associated with contact loss, dendrite formation, and chemo-mechanical instability \cite{Xia2019ChemPR}. The integration of mechanical, chemical, and artificial SEI engineering strategies remains central to stabilizing these interfaces and enabling reliable long-term operation.

Collectively, these studies illustrate that a mechanistic understanding of SEI formation, evolution, and control is essential for bridging current lithium-ion technologies with next-generation high-energy, fast-charging, and solid-state energy storage systems for electric vehicles.

\begin{table*}[!htbp]
\centering
\small
\caption{Key References on SEI in Lithium-Ion Batteries}
\label{tab:sei_subnode}
\begin{tabular}{p{3.5cm} c p{6.5cm} p{5.5cm}}
\hline
\textbf{Authors} & \textbf{Year} & \textbf{Key Contributions} & \textbf{Limitations} \\
\hline

Zhang \textit{et al.} \cite{Zhang2025ArtSEI} & 2025 &
Summarized advances in artificial SEI (Art-SEI) fabrication, emphasizing tailored structures and compositions to enhance stability and ionic conductivity. &
Primarily review; experimental validation across diverse battery chemistries limited. \\

Liang \textit{et al.} \cite{Liang2025DualHalideSEI} & 2025 &
Developed highly ductile dual-halide SEI layers on lithium metal anodes, improving Li$^+$ transport, suppressing dendrites, and enhancing cycling stability. &
Focused on Li-metal anodes; application to full cells and commercial-scale systems not fully demonstrated. \\

Hao \textit{et al.} \cite{Hao2025SEIStructure} & 2025 &
Provided comprehensive analysis of SEI structural evolution at electrode-electrolyte interfaces, detailing formation mechanisms and degradation pathways. &
Review-based; experimental quantification across different electrodes limited. \\

Ryu \textit{et al.} \cite{Ryu2025VibrationSEI} & 2025 &
Demonstrated vibration-induced SEI thinning and inorganic enrichment, enhancing lithium-ion transport and battery performance under dynamic conditions. &
Specific to mechanical vibration conditions; generalizability to standard EV operation requires further study. \\

Li \textit{et al.} \cite{Li2024SEIMechanical} & 2024 &
Reviewed mechanical stability of SEI, highlighting stress accumulation, fracture, and strategies to improve durability via in situ/ex situ modifications. &
Focus on mechanical aspects; chemical and electrochemical interactions less emphasized. \\

Adenusi \textit{et al.} \cite{Adenusi2023AEM} & 2023 &
Reviewed SEI engineering via electrolyte additives and surface modifications to stabilize electrode–electrolyte interfaces. &
Mainly graphite anodes; lacks comprehensive coverage of solid-state and lithium metal systems. \\

Beheshti \textit{et al.} \cite{Beheshti2022iScience} & 2022 &
Systematic examination of SEI formation, retention, and additive effects on graphite electrodes, linking to long-term electrochemical performance. &
Focused on graphite; emerging chemistries and full cell integration not fully addressed. \\

Lin \textit{et al.} \cite{Lin2020NanoEnergy} & 2020 &
Analyzed interfacial phenomena in SEI evolution and strategies for enhancing stability under cycling. &
Limited consideration of high-power fast-charging and temperature effects. \\

Li \textit{et al.} \cite{Li2020CRPS} & 2020 &
Analyzed fast charging effects, lithium plating, and SEI degradation in EV relevant lithium-ion batteries. &
Specific to Li-ion; solid-state and metal-anode architectures only briefly discussed. \\

\hline
\end{tabular}
\end{table*}

\subsection{\textbf{Beyond Lithium-Ion: Sodium, Metal-Ion, and Metal-Air Batteries}}
Although lithium-ion batteries currently dominate the EV market, their dependence on relatively scarce and geopolitically sensitive resources such as cobalt and nickel, together with intrinsic limitations in energy density and safety, has intensified research into alternative battery chemistries. In this context, sodium-ion, sodium-beta alumina, metal-ion, and metal-air batteries have emerged as promising candidates for both stationary and mobile energy storage, offering advantages in terms of resource abundance, cost reduction, and potential energy density \cite{Asghar2021,Solyali2022}.

Sodium-based batteries exploit the high natural abundance and low cost of sodium compared to lithium, making them attractive for large scale and cost sensitive applications. Among these technologies, sodium-beta alumina batteries based on Na-\(\beta\)Al\(_2\)O\(_3\) solid electrolytes exhibit high ionic conductivity, excellent thermal stability, and favorable cycle life at elevated operating temperatures \cite{Yang2022Sodium,Fertig2022}. Numerical investigations have highlighted the critical role of thermal management in such systems, demonstrating that optimized natural convection and heat transfer in vertical cell configurations with lateral baffles are essential to avoid local hotspots and ensure stable operation \cite{Yang2022Sodium}.

In parallel, room temperature sodium-metal and sodium-sulfur batteries represent a significant advancement toward EV relevant sodium-based systems. However, their practical implementation is constrained by interfacial instability at the sodium-metal anode, which can induce dendrite formation, low Coulombic efficiency, and rapid capacity fade \cite{Soni2022,Zhu2021}. To address these challenges, strategies such as in situ formation of stable interphase layers, controlled electrolyte composition, and the use of polymer based separators have demonstrated improved interfacial stability and cycling performance \cite{Zhu2021,Eng2021,Jeon2021}. Polymer electrolytes, in particular, offer enhanced safety, mechanical flexibility, and extended cycle life, making them attractive for future EV integration \cite{Au2022}.

Recent developments in sodium based batteries have also focused on cathode engineering, particularly layered sodium transition metal oxides. Optimization of electrode microstructure, particle morphology, and phase stability has been shown to improve ionic diffusion kinetics and mechanical robustness, which are critical for achieving high rate capability and long cycle life \cite{Fertig2022}.

Beyond sodium-based chemistries, metal-ion batteries such as zinc-ion, potassium-ion, and magnesium-ion systems have attracted increasing attention due to their low cost, environmental compatibility, and inherent safety, particularly in aqueous configurations \cite{Verma2021}. Significant progress has been achieved through the development of advanced cathode materials, including vanadium pentoxide, manganese oxides, and conductive polymer composites, which enhance interlayer spacing, facilitate ion transport, and maintain structural integrity during cycling \cite{Zhang2021Zn,Yang2022Zn,Wang2022Zn}.

Despite these advances, metal-ion batteries face several persistent challenges, including limited operating voltage windows, dendrite formation, and electrolyte decomposition. Recent studies have demonstrated that advanced electrode architectures, conductive polymer coatings, and engineered interfacial layers can effectively mitigate these issues, enabling improved rate performance and mechanical flexibility suitable for portable electronics and emerging EV-related applications \cite{Zhang2021Zn,Yang2022Zn,Wang2022Zn}.

Metal-air batteries, including zinc-air, lithium-air, and flexible metal-air systems, offer some of the highest theoretical energy densities among electrochemical energy storage technologies \cite{Olabi2021,Peng2021,Wang2022MetalAir,Lee2011,Leong2022,Abraham2008}. By utilizing oxygen from ambient air as the cathode reactant, these systems significantly reduce active material mass and enable ultra-high energy density configurations.

However, practical deployment of metal-air batteries is hindered by challenges related to catalyst stability, oxygen diffusion, electrolyte management, and air cathode degradation. Research efforts have focused on both aqueous and non-aqueous electrolyte systems, the development of advanced bifunctional catalysts, and protective coatings to enhance cycling stability and energy efficiency \cite{Wang2022MetalAir,Olabi2021,Lee2011}. Flexible metal-air batteries have been explored primarily for wearable and portable electronics, where mechanical robustness, bending tolerance, and operational stability are critical performance metrics \cite{Peng2021}. Non aqueous lithium-air batteries, in particular, aim to maximize energy density but require highly controlled environments to suppress parasitic reactions and electrolyte decomposition \cite{Abraham2008}.

Future research directions across non lithium battery chemistries increasingly emphasize hybrid systems, solid state electrolytes, interphase engineering, and scalable manufacturing processes to bridge the performance gap between lithium-ion batteries and next-generation EV energy storage technologies \cite{Asghar2021,Solyali2022,Linden2015}. These strategies aim to deliver safe, high energy density, and cost effective batteries capable of supporting large-scale EV deployment while reducing environmental impact and supply-chain constraints.

\begin{table*}[!htbp]
\centering
\small
\caption{Key References on Beyond Lithium-Ion Batteries (Sodium, Metal-Ion, Metal-Air) for EVs}
\label{tab:beyond_li_ion}
\begin{tabular}{p{3.5cm} c p{6.5cm} p{5.5cm}}
\hline
\textbf{Authors} & \textbf{Year} & \textbf{Key Contributions} & \textbf{Limitations} \\
\hline

Weilong Liu \textit{et al.} \cite{Liu2026} & 2026 &
Investigated 2D Me-C8B5 monolayer as high-performance anode for Na/K-ion batteries using DFT; demonstrated high capacity, low diffusion barriers, and excellent thermal stability. &
Computational study only; experimental validation needed. \\

Jan Koloch \textit{et al.} \cite{Koloch2025} & 2025 &
Techno-economic analysis of sodium-ion batteries (SIBs) for EVs; SIBs show lower cost per km compared to NMC/LFP and enable higher max range with low-capacity vehicles. &
Packaging constraints limit high-capacity SIB EVs; slight reduction in max range vs Li-ion. \\

Liu Pei \textit{et al.} \cite{Liu2025} & 2025 &
Developed solid-state sodium-ion batteries with closed-loop feedback-optimized composite electrolytes; high conductivity and 90\% capacity retention after 700 cycles. &
Focus on lab-scale cells; scale-up and long-term EV performance not assessed. \\

Yao \textit{et al.} \cite{Yao2025} & 2025 &
Assessed sodium-ion technology roadmaps vs Li-ion; identified pathways for potential cost competitiveness in 2030s and highlighted sensitivity to mineral supply chains. &
Price competitiveness challenging in near term; heavily model-dependent assumptions. \\

Quansheng Li \textit{et al.} \cite{Li2025} & 2025 &
Designed laminated 3D network Li4Ti5O12 electrodes via femtosecond laser; improved volumetric capacity, rate capability, and cycling stability for metal-ion batteries. &
Experimental work focused on lab-scale anodes; EV-scale application not demonstrated. \\

Muhammad Akbar \textit{et al.} \cite{Akbar2025} & 2025 &
DFT study of SiB2 monolayer as anode for Li/Na/K-ion batteries; high storage capacity, metallic conductivity, and stability during lithiation/sodiation/potassiation. &
Theoretical study only; synthesis and real battery tests not performed. \\

Shabeer \textit{et al.} \cite{Shabeer2025} & 2025 &
Reviewed metal-air batteries as EV range extenders; analyzed energy density, cost, and performance trade-offs for different metal-air chemistries. &
Primarily review; practical implementation and cycling stability data limited. \\

Chang Guo \textit{et al.} \cite{Guo2025} & 2025 &
Developed light-assisted low-pressure Li-CO2 metal-air batteries; showed improved catalytic performance and reaction kinetics under extreme low-pressure conditions. &
Focused on theoretical and lab-scale validation; real-world EV performance untested. \\

S. S. Shinde \textit{et al.} \cite{Shinde2025} & 2025 &
Outlined design strategies for practical zinc-air batteries (ZABs) for EVs; discussed interface dynamics, scale-up, and testing protocols for Ah-scale cells. &
Perspective article; practical large-scale testing and long-term cycling data limited. \\

\hline
\end{tabular}
\end{table*}

\section{\textbf{Next-Generation Intelligent Batteries for Electric Vehicles}}
\subsection{\textbf{Machine Learning Applications in Battery Systems}}

\begin{figure*}[!htbp]
\centering
\includegraphics[width=\textwidth, height=0.7\textheight, keepaspectratio]{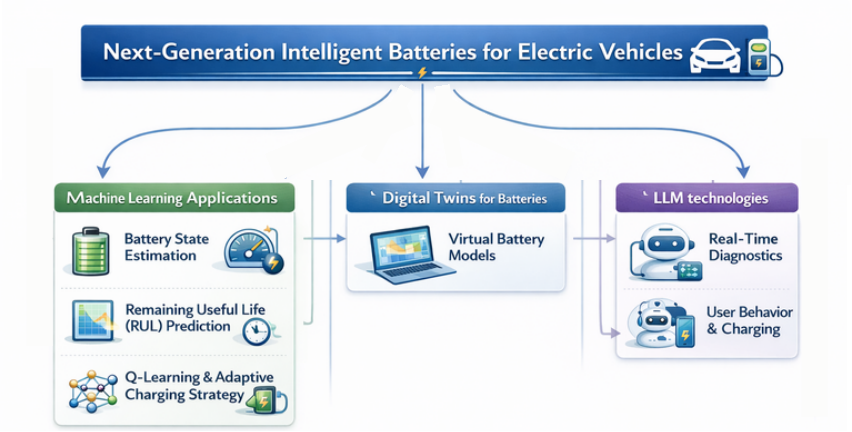}
\caption{Taxonomy of next-generation intelligent batteries for electric vehicles}
\label{fig:taxonomy_batteries}
\end{figure*}

The rapid electrification of transportation has fundamentally transformed the role of lithium-ion batteries from passive energy storage devices into intelligent, cyber physical systems. While conventional battery research has historically focused on materials, electrochemistry, and thermal management, the increasing complexity of EV operation now necessitates data-driven intelligence for real-time monitoring, prediction, optimization, and decision-making. AI has consequently emerged as a central enabler for next-generation intelligent battery systems, enhancing safety, performance, longevity, and sustainability across the entire battery lifecycle.

Modern BMS increasingly rely on AI to overcome the limitations of physics based and equivalent circuit models under dynamic operating conditions. A major paradigm shift is the introduction of digital twins, which combine real time sensor data with machine learning models to create adaptive virtual replicas of battery systems. Pooyandeh and Sohn proposed an AI-empowered digital twin framework based on time series generative adversarial networks, capable of synthesizing accurate state of charge trajectories and improving energy efficiency, operational safety, and battery lifespan in EV applications \cite{Pooyandeh2023Mathematics}.

Complementary AI-driven BMS architectures have integrated advanced sensing modalities, such as X-ray computed tomography and electrochemical impedance spectroscopy, to enable model free estimation of state of charge, state of health, and fault dynamics \cite{Palanichamy2024EATP}. These approaches significantly enhance fault detection sensitivity and thermal risk mitigation, highlighting the value of combining multimodal data with machine learning inference. At the system level, cloud connected and IoT enabled BMS frameworks further support adaptive diagnostics, distributed thermal management, and long term reliability under real-world operating conditions \cite{Farman2024E3S}.

Accurate estimation of battery states and RUL is critical for ensuring EV safety, performance optimization, and cost efficiency. AI-based filtering and learning algorithms have demonstrated superior robustness compared to traditional observers under noisy and nonlinear conditions. Ahwiadi and Wang introduced an AI-driven particle filter with adaptive degeneracy detection, achieving substantial reductions in state estimation and RUL prediction errors while maintaining computational efficiency suitable for real time BMS deployment \cite{Ahwiadi2024Batteries}.

ML \cite{Bahi2023Homomorphic,Bahi2025SFNN} regression and ensemble techniques have further advanced battery prognostics under variable load and temperature profiles. Models based on gradient boosting, bagging regressors, and support vector machines have shown high predictive accuracy and generalization capability for RUL estimation \cite{Sravanthi2025JETIA}. Deep transfer learning approaches have also been explored to capture degradation dynamics across diverse operating conditions, demonstrating the potential of feature reuse and cross-domain learning in battery aging prediction \cite{Zhang2024Batteries}.

Recent progress in time series modeling has reinforced the role of recurrent neural networks in battery health forecasting. Optimized bidirectional LSTM architectures trained with metaheuristic optimizers have achieved high fidelity capacity fade and SOH predictions, benefiting from improved global local parameter balance during training \cite{Wang2025EvolIntell}. Hybrid DL architectures combining convolutional, GRU, and LSTM layers further enable accurate and computationally efficient state estimation suitable for onboard implementation \cite{Shahriar2022Energies,Noreen2024Network}.

Beyond passive monitoring, AI enables active optimization of battery operation through RL \cite{Bahi2023MovieRec,Bahi2025DroneEnergy,Bahi2025DataCenterDRL} and adaptive control. Q-learning based BMS optimization strategies have demonstrated measurable improvements in energy efficiency and battery lifespan by dynamically adjusting operating policies in response to usage patterns and environmental conditions \cite{Suanpang2024Sustainability}. Such approaches move battery control from rule-based logic toward self learning systems capable of balancing performance, safety, and degradation in real time.

At the energy management level, AI-integrated control frameworks have been shown to enhance EV range, predictive maintenance, and adaptive route planning through coordinated optimization of battery usage and vehicle dynamics \cite{Arevalo2024WEVJ}. However, these advances also expose challenges related to cybersecurity, interoperability, and lightweight algorithm design, particularly for real time safety indicators such as state of power and state of energy monitoring \cite{Ghazali2025Symmetry,Challoob2024EEE}.

The influence of AI on next generation batteries extends well beyond operational BMS functions. ML is increasingly transforming the battery value chain, from materials discovery to end of life management. AI-driven workflows have accelerated the identification of high energy density electrode materials and optimized electrolyte formulations, reducing reliance on costly trial and error experimentation \cite{Acharya2024Elsevier,Feng2024AJMC,Alzamer2025Crystals}.

Explainable AI models have further contributed to performance optimization by revealing key degradation drivers, such as temperature sensitivity, thereby improving transparency and trust in data driven decision making \cite{Oyucu2024Sustainability}. In manufacturing, AI-enabled automation and quality control enhance process consistency, while blockchain supported traceability frameworks improve sustainability and recycling efficiency across the battery lifecycle \cite{Acharya2024Elsevier}.

At the system scale, AI enhanced energy storage systems play a critical role in enabling renewable energy integration, grid flexibility, and emission reduction. The synergy between battery storage, AI optimization, and supportive policy frameworks has been identified as a key factor for large scale adoption of sustainable energy systems \cite{Razmjoo2024Sustainability}. Hybrid architectures combining lithium-ion batteries with supercapacitors further illustrate the potential of intelligent power management to improve thermal stability, internal resistance estimation, and system longevity, while highlighting the need for advanced control strategies and scalable modeling approaches \cite{Challoob2024EEE}.

Overall, AI-driven intelligent lithium-ion batteries represent a fundamental shift in how energy storage is designed, managed, and deployed in electric vehicles. By unifying data driven intelligence with electrochemical insight, next generation lithium-ion batteries evolve into adaptive, predictive, and self-optimizing systems. Continued progress in AI algorithms, sensor integration, and system level co-design will be essential for realizing safe, high performance, and environmentally sustainable electric mobility.
\begin{table*}[!htbp]
\centering
\small
\caption{Key References on AI for Battery Management and Energy Storage Systems in EV}
\label{tab:ai_bms}
\begin{tabular}{p{3.5cm} c p{6.5cm} p{5.5cm}}
\hline
\textbf{Authors} & \textbf{Year} & \textbf{Key Contributions} & \textbf{Limitations} \\
\hline

Wang \textit{et al.} \cite{Wang2025EvolIntell} & 2025 &
Bi-directional LSTM model optimized using an adaptive Gold Rush Optimizer for lithium-ion battery capacity prediction. & Model complexity and real-time deployment constraints were not fully addressed. \\

Ghazali \textit{et al.} \cite{Ghazali2025Symmetry} & 2025 &
Provided a comprehensive review of advanced algorithms used in EV battery management systems, covering estimation, control, and optimization techniques. &
Primarily qualitative analysis; lacks experimental benchmarking across different BMS architectures. \\

Alzamer \textit{et al.} \cite{Alzamer2025Crystals} & 2025 &
Highlighting data driven acceleration of Li-ion battery innovation. &
Limited discussion on system level integration in EVs. \\

Sravanthi \textit{et al.} \cite{Sravanthi2025JETIA} & 2025 &
Evaluated multiple machine learning regression models for remaining useful life prediction of lithium-ion batteries in EVs. &
Relies on historical datasets; robustness under real world operating variability remains unclear. \\

Madhavan \textit{et al.} \cite{Madhavan2025IJHE} & 2025 &
Applied LSTM based time series modeling for monitoring pressure valves in hydrogen-powered vehicles and infrastructure. &
Not directly focused on battery systems; applicability to EV BMS is indirect. \\

Palanichamy \textit{et al.} \cite{Palanichamy2024EATP} & 2024 &
Presented AI-driven optimization strategies for battery management systems to enhance EV performance and operational efficiency. &
Lacks detailed validation on large scale or real world EV datasets. \\

Farman \textit{et al.} \cite{Farman2024E3S} & 2024 &
Proposed an AI-enhanced BMS framework emphasizing safety improvement, performance optimization, and battery longevity. &
Conceptual framework with limited quantitative performance comparison. \\

Ahwiadi \textit{et al.} \cite{Ahwiadi2024Batteries} & 2024 &
Introduced an AI-driven particle filter approach for accurate battery state estimation and RUL prediction. &
Computational cost and scalability for real time EV applications were not fully analyzed. \\

Suanpang \textit{et al.} \cite{Suanpang2024Sustainability} & 2024 &
Developed a Q-learning based battery management strategy to improve sustainability and energy efficiency in EVs. &
Convergence speed and adaptability to dynamic driving conditions require further investigation. \\

Miraftabzadeh \textit{et al.} \cite{Miraftabzadeh2024Electronics} & 2024 &
Explored the synergy between artificial intelligence and energy storage systems for EV applications. &
Broad scope review with limited depth on specific BMS algorithms. \\

Badran \textit{et al.} \cite{Badran2024Pertanika} & 2024 &
Analyzed AI techniques applied to EV battery management systems, highlighting challenges and open issues. &
Does not propose a concrete implementation or validation framework. \\

Oyucu \textit{et al.} \cite{Oyucu2024Sustainability} & 2024 &
Integrated machine learning with explainable AI to enhance lithium-ion battery energy management and transparency. &
Explainability methods may introduce additional computational overhead. \\

Feng \cite{Feng2024AJMC} & 2024 &
Discussed AI-based transformation of lithium battery material design and optimization processes. &
Mainly conceptual; lacks experimental or industrial validation. \\

Noreen \textit{et al.} \cite{Noreen2024Network} & 2024 &
Compared traditional and modern battery state estimation methods for EV battery management systems. &
Limited focus on advanced deep learning or hybrid AI models. \\

Arevalo \textit{et al.} \cite{Arevalo2024WEVJ} & 2024 &
Provided a systematic review of AI integration into energy management systems for electric vehicles. &
Survey oriented study without performance benchmarking. \\

Challoob \textit{et al.} \cite{Challoob2024EEE} & 2024 &
Reviewed energy and battery management systems for EVs and proposed practical recommendations. &
Recommendations are qualitative and require experimental validation. \\

Zhang \textit{et al.} \cite{Zhang2024Batteries} & 2024 &
Proposed a deep transfer learning framework for lithium-ion battery life prediction. &
Transferability across battery chemistries and usage patterns was not fully assessed. \\

Razmjoo \textit{et al.} \cite{Razmjoo2024Sustainability} & 2024 &
Investigated AI-assisted expansion of energy storage systems within renewable energy frameworks. &
Focuses more on grid-level systems than EV specific battery management. \\

Pooyandeh \textit{et al.} \cite{Pooyandeh2023Mathematics} & 2023 &
Introduced an AI-empowered digital twin framework for real time monitoring of lithium-ion batteries in EVs. &
Digital twin accuracy depends heavily on data availability and model calibration. \\

\hline
\end{tabular}
\end{table*}

\subsection{\textbf{Digital Twin Applications in Battery Systems}}

DT technology is rapidly emerging as a transformative approach for lithium-ion battery management, bridging physical systems with high fidelity virtual representations to enable real time monitoring, predictive maintenance, and lifecycle optimization. By integrating advanced sensing, AI-driven analytics, and interactive visualization, DTs provide a robust framework for enhancing battery performance, safety, and longevity across EV applications.

Digital twins for lithium-ion batteries rely on tight synchronization between physical and virtual systems, incorporating real time sensor data, model-based estimators, and predictive algorithms. Wang and Li demonstrated a hybrid system combining EKF for SOC estimation and PSO for SOH prediction. This integrated approach effectively mitigated the impact of Gaussian noise in real time scenarios, validating the practical feasibility of DT based battery prognostics in embedded platforms \cite{Zhou2021ICIVIS}. Similarly, Kang et al. illustrated the utility of DTs in hybrid powertrains combining lithium-ion batteries and fuel cells, emphasizing three dimensional visualization and mathematical modeling to enable dynamic energy control and scalable simulation \cite{Kang2024Batteries}.

Accurate SOC and SOH estimation is central to DT functionality. Tang et al. proposed a HIF-PF within a DT framework, achieving SOC estimation errors as low as 0.14\% under dynamic stress tests. The system utilized interactive visualization to facilitate real time monitoring and supports future closed-loop optimization of physical battery parameters \cite{Tang2022JES}. Song et al. introduced a CNN-LSTM hybrid model to capture both spatial and temporal dependencies in battery signals, demonstrating robust SOC prediction with mean absolute errors below 1.5\% across variable temperatures. A periodic parameter update mechanism enhanced adaptability to battery aging, supporting practical deployment with minimal recalibration \cite{Song2019IEEE}. 

Khalid and Sarwat further explored unified machine learning architectures integrating ARIMA, NARX, and MLP networks with Akaike Information Criterion (AIC) optimization, achieving RMSE as low as 0.1323\% under low C-rate conditions. While primarily offline, these frameworks indicate the potential for real-time SOC forecasting in aging-aware, multi pack DTenabled systems \cite{Khalid2021IEEE}. Zhao et al. advanced hybrid LSTM-EKF DT architectures, combining accurate initial SOC estimation with EKF-based online correction to improve robustness, reduce error margins, and enable hierarchical DT models capable of simultaneous SOC and RUL prediction \cite{Zhao2022ENERGYCON}.

The fusion of DTs with AI and IoT technologies enables enhanced real time diagnostics, anomaly detection, and system optimization. Lakshmi and Sarma presented a DT framework incorporating AI and IoT for EV energy storage, achieving low latency (0.12 s) and fast feedback (0.45 s), while improving energy efficiency by 15\% and minimizing capacity fade to 0.0025\% per cycle. Their system achieved high anomaly detection precision (95\%) and detection rate (76\%), demonstrating the practical efficacy of AI-IoT-DT integration for intelligent battery lifecycle management \cite{Lakshmi2025ICMSCI}. 

Beyond vehicle-level applications, DTs are increasingly leveraged in energy and smart grid management. Das et al. reviewed the integration of DTs with machine learning in power systems, highlighting applications in battery health prognosis, renewable energy integration, and cost forecasting, demonstrating the broad scalability and versatility of DT enabled analytics \cite{Das2024EnergyX}. Kang emphasized edge cloud DT architectures and AutoML for smart vehicles, showing efficient deployment of computationally intensive AI analytics under resource constraints, crucial for autonomous driving and intelligent mobility systems \cite{Kang2025JWE}. 

From a systems engineering standpoint, DTs enhance lifecycle management by enabling hierarchical monitoring, predictive maintenance, and adaptive control strategies. Chai et al. highlighted DT deployment in intelligent connected vehicles (ICVs), emphasizing modular data handling, partitioning between edge and cloud infrastructure, and lifecycle optimization \cite{Chai2024Elektron}. Wang et al. further underscored the potential of DT frameworks in energy system operations, enabling real time simulation, predictive analytics, and system wide optimization for smart energy ecosystems \cite{Wang2022GreenEnergy}. Collectively, these studies confirm that DTs provide a unified platform for battery SOC/SOH estimation, RUL prediction, energy management, and integration with broader EV and smart grid operations.

Digital twin technology represents a paradigm shift in EV battery management, moving from reactive monitoring toward predictive, adaptive, and autonomous optimization. By integrating AI, IoT, and edge cloud capabilities, DTs facilitate robust SOC/SOH estimation, anomaly detection, and RUL forecasting, while supporting system level decisions in vehicle and grid contexts. Future research directions include hierarchical DT modeling for multi cell and multi pack systems, adaptive learning under evolving battery aging patterns, and tighter integration with V2G and renewable energy frameworks, ultimately enhancing battery reliability, safety, and environmental sustainability.

\begin{table*}[!htbp]
\centering
\small
\caption{Key References for DT in Battery Systems}
\label{tab:digital_twin_bms}
\begin{tabular}{p{3.5cm} c p{6.5cm} p{5.5cm}}
\hline
\textbf{Authors} & \textbf{Year} & \textbf{Key Contributions} & \textbf{Limitations} \\
\hline

Kang \textit{et al.} \cite{Kang2025JWE} & 2025 &
Presented software practice and experience on smart mobility digital twins for transportation and automotive industry, highlighting EV edge cloud integration and AutoML techniques. &
Focuses on software platforms; limited experimental validation in real EV deployments. \\

Lakshmi \textit{et al.} \cite{Lakshmi2025ICMSCI} & 2025 &
Developed energy storage systems using digital twins with AI and IoT for efficient energy management and prolonged battery life in electric vehicles. &
Primarily simulation based; scalability to large fleets not fully addressed. \\

Kang \textit{et al.} \cite{Kang2024Batteries} & 2024 &
Proposed digital twin enhanced control for fuel cell and lithium-ion battery hybrid vehicles, improving system performance and efficiency. &
Limited demonstration on real world hybrid EVs; integration challenges remain. \\

Das \textit{et al.} \cite{Das2024EnergyX} & 2024 &
Comprehensive review of digital twin technology and machine learning applications in energy systems, smart grids, renewable energy, and EV optimisation. &
Review-focused; lacks novel experimental contributions. \\

Chai \textit{et al.} \cite{Chai2024Elektron} & 2024 &
Surveyed recent progress on digital twins in intelligent connected vehicles, discussing architectures, methodologies, and automotive applications. &
Emphasizes overview; detailed case studies and benchmarks are limited. \\

Tang \textit{et al.} \cite{Tang2022JES} & 2022 &
Designed lithium battery management systems based on digital twin frameworks, improving SOC estimation and system reliability. &
Limited validation in high power EV systems; mainly lab scale implementation. \\

Wang \textit{et al.} \cite{Wang2022GreenEnergy} & 2022 &
Reviewed digital twin techniques in smart manufacturing and energy management applications, including batteries and EVs. &
Focuses on surveys; experimental or industrial validation limited. \\

Zhao \textit{et al.} \cite{Zhao2022ENERGYCON} & 2022 &
Developed digital twin driven estimation methods for SOC in Li-ion batteries, enhancing prediction accuracy and monitoring. &
Tested on specific battery models; generalization to other chemistries not demonstrated. \\

Khalid \textit{et al.} \cite{Khalid2021IEEE} & 2021 &
Introduced univariate neural network models for SOC forecasting using minimized Akaike Information Criterion for Li-ion batteries. &
Limited to SOC prediction; integration with full digital twin frameworks not explored. \\

Zhou \textit{et al.} \cite{Zhou2021ICIVIS} & 2021 &
Presented a digital twin model for battery management systems, covering concepts, algorithms, and platform implementation. &
Conceptual and methodological; experimental deployment in EVs not fully addressed. \\

Song \textit{et al.} \cite{Song2019IEEE} & 2019 &
Proposed combined CNN-LSTM network for SOC estimation of Li-ion batteries. &
Focuses on machine learning model; not a full digital twin implementation. \\

\hline
\end{tabular}
\end{table*}

\subsection{\textbf{Large Language Models Applications in Battery Systems}}

LLMs have emerged as a transformative tool in the management and optimization of EV battery systems, leveraging advanced natural language understanding, multimodal reasoning, and generative capabilities to enhance prediction, control, and user interaction. The integration of LLMs enables sophisticated analysis of high-dimensional battery data, predictive maintenance, adaptive scheduling, and real time decision-making, bridging the gap between raw telemetry and actionable insights.

Honnalli and Farooq \cite{Honnalli2025COINS} demonstrated the potential of multimodal LLM guided frameworks for securing EV charging infrastructure. By combining sequential time-series SOC forecasting with LLM based visual reasoning, their approach detects cyber threats such as overcharging, denial-of-service attacks, and billing fraud, significantly improving early-stage anomaly detection and reducing energy losses. Similarly, Honnalli and Farooq \cite{Honnalli2025WoWMoM} extended this methodology using RAG to integrate domain knowledge with live charging session data, enhancing classification accuracy and operational resilience against evolving cyber attacks.

Abraham et al. \cite{Abraham2025Springer} explored LLM applications in broader intelligent transportation systems, emphasizing sustainable EV integration. They highlighted how LLMs optimize battery use, charging infrastructure, energy management, and predictive maintenance while supporting autonomous driving and smart city initiatives. Generative AI aids in route optimization, emissions reduction, and public education, while MLLMs enable simultaneous interpretation of textual, numerical, and visual battery related data.

For precise battery state estimation, Yunusoglu et al. \cite{Yunusoglu2025ISQED} introduced a transformer based LLM framework to predict the SOH and RUL of lithium titanate cells. The model, trained on cycle-based and instantaneous discharge data, achieved a mean absolute error of 0.87\% and successfully detected early degradation signs, enabling real-time predictive maintenance and improving energy efficiency. Complementing this, Tuncel et al. \cite{Tuncel2025EnergyReports} developed an LLM driven pipeline that automates the machine learning workflow for SOH prediction, including data preprocessing, feature selection, model recommendation, and hyperparameter tuning, achieving a 52\% reduction in MAPE compared to traditional methods.

Advanced multimodal reasoning capabilities are illustrated by Zhang et al. \cite{Zhang2026Energy}, who combined LLMs with a MSIformer for LIB SOH estimation. By fusing textual embeddings derived from contextual prompts with numerical battery features, their model captured multi timescale degradation dynamics, improving robustness under extreme operating conditions and achieving substantial reductions in RMSE, MAE, and MAPE relative to nine benchmarks. Similarly, Liu et al. \cite{Liu2025arxiv} applied expert guided LLM reasoning to battery material discovery, demonstrating the identification, synthesis, and characterization of novel cathode materials with practical capacity improvements, highlighting the potential of LLMs for accelerating materials innovation in battery design.

LLMs also facilitate user-centric EV operations and grid interactions. Sun et al. \cite{Sun2025arxiv} proposed an LLM driven multi agent framework to construct user digital twins for smart charging stations, optimizing vehicle to grid interactions through dynamic incentive mechanisms and predictive simulation of user behavior. Ji et al. \cite{Ji2025CEEGE} and Feng et al. \cite{Feng2025ACPEE} further demonstrated LLM-based frameworks to synthesize EV charging databases and simulate user behavior, enabling precise load forecasting, adaptive scheduling, and scenario-based planning for urban EV networks.

In the context of operational optimization, Zafar and Bayhan \cite{Zafar2025CPE} leveraged LLMs to solve the EVRP with advanced prompting, reducing range anxiety and efficiently incorporating charging station constraints into routing. Lozano and Shi \cite{Lozano2025NeurIPS} applied LLM agents to optimize the dispatch of mobile chargers for construction EVs, translating high-level objectives into solver ready inputs and ensuring feasible, cost efficient, and environmentally optimized operations. Yang et al. \cite{Yang2025CCSSTA} extended LLM applications to battery health aware scheduling in building energy systems, demonstrating interpretable SOH based strategies for lifespan extension and adaptive control in energy storage applications.

Finally, Zhou et al. \cite{Zhou2025SETA} presented BatteryGPT, a GPT-based framework for lithium-ion battery charging state prediction in swapping stations. By fine tuning only the input embedding and output layers and using contrastive temporal embeddings, the model achieved a 55.52\% improvement in prediction accuracy over conventional LSTM based methods, showcasing the scalability and efficiency of LLM driven predictive control in industrial EV energy management.

These studies illustrate the multifaceted applications of LLMs in battery systems, encompassing predictive maintenance, state estimation, energy scheduling, cybersecurity, user behavior modeling, routing optimization, fault diagnosis, demand forecasting, anomaly detection, and material discovery. The semantic reasoning, multimodal integration, adaptive capabilities, contextual awareness, and real-time decision-making of LLMs provide a unified platform for intelligent battery management, supporting next-generation EV ecosystems with enhanced efficiency, reliability, scalability, and sustainability.

\begin{table*}[!htbp]
\centering
\small
\caption{Key References for LLM-Driven Battery and EV Systems}
\label{tab:llm_ev_battery}
\begin{tabular}{p{3.5cm} c p{6.5cm} p{5.5cm}}
\hline
\textbf{Authors} & \textbf{Year} & \textbf{Key Contributions} & \textbf{Limitations} \\
\hline

Zhang \textit{et al.} \cite{Zhang2026Energy} & 2026 &
Proposed LLM-MSIformer, combining large language models and multi-time scale interval transformers for Li-ion battery SOH estimation. &
Primarily simulation-based; real-world deployment not demonstrated. \\

Honnalli \textit{et al.} \cite{Honnalli2025COINS} & 2025 &
Developed a multimodal LLM-guided sequential approach for detecting cyber threats in EV charging systems. &
Limited to EV charging cybersecurity; broader energy systems not addressed. \\

Yunusoglu \textit{et al.} \cite{Yunusoglu2025ISQED} & 2025 &
Introduced an LLM framework for Li-ion battery state-of-health estimation. &
Validation mostly on benchmark datasets; real-world variability may affect performance. \\

Tuncel \textit{et al.} \cite{Tuncel2025EnergyReports} & 2025 &
LLM-based methodology for data-driven health prediction of Li-ion batteries. &
Focus on data-driven models; physical degradation mechanisms less emphasized. \\

Liu \textit{et al.} \cite{Liu2025arxiv} & 2025 &
Expert-guided LLM reasoning for battery discovery from hypothesis generation to synthesis and characterization. &
Lab-scale demonstrations; large-scale production integration not covered. \\

Sun \textit{et al.} \cite{Sun2025arxiv} & 2025 &
LLM-driven user digital twin for dynamic incentive strategies in smart EV charging stations. &
Conceptual and simulation-focused; lacks large-scale deployment evaluation. \\

Ji \textit{et al.} \cite{Ji2025CEEGE} & 2025 &
Applied LLM-generated data to analyze EV user charging characteristics. &
Limited generalization to diverse EV populations and geographies. \\

Feng \textit{et al.} \cite{Feng2025ACPEE} & 2025 &
LLM-based agent framework for simulating EV charging behavior. &
Simulation-focused; real-world validation is missing. \\

Zafar \textit{et al.} \cite{Zafar2025CPE} & 2025 &
Used LLMs to solve the EV routing problem with advanced prompting techniques. &
Focus on routing algorithms; energy management and real-world uncertainties not fully included. \\

Honnalli \textit{et al.} \cite{Honnalli2025WoWMoM} & 2025 &
LLM-powered agentic AI approach to secure EV charging systems against cyber threats. &
Limited scope to charging infrastructure; broader vehicle network security not included. \\

Zhou \textit{et al.} \cite{Zhou2025SETA} & 2025 &
Applied GPT models for predicting Li-ion battery charging state in battery swapping stations. &
Focused on swapping stations; generalization to other battery types not addressed. \\

Lozano \textit{et al.} \cite{Lozano2025NeurIPS} & 2025 &
LLM agent for dispatching mobile chargers in microgrids, enabling optimization of EV charging. &
Demonstrated in microgrid scenarios; industrial-scale feasibility not validated. \\

Yang \textit{et al.} \cite{Yang2025CCSSTA} & 2025 &
LLM-based battery health-aware scheduling (LLM-BAS) in building energy systems. &
Limited to building energy systems; broader transportation applications not explored. \\

Abraham \textit{et al.} \cite{Abraham2025Springer} & 2025 &
Harnessed LLMs for sustainable and intelligent transportation systems in the EV era. &
Conceptual and review-focused; experimental validation limited. \\

\hline
\end{tabular}
\end{table*}

\section{\textbf{Discussion of Open Challenges, Research Gaps, and Future Directions}}

Despite significant advances in electrochemical energy storage and the emergence of intelligent battery technologies, several open challenges and research gaps remain, which need to be addressed to fully realize the potential of next-generation battery systems for EVs.

Current lithium-ion batteries face intrinsic limitations in energy density, cycle life, and safety. While alternative chemistries such as sodium-ion, metal-ion, and metal-air batteries offer promising avenues, challenges such as low Coulombic efficiency, poor rate capability, dendrite formation, and limited cycle stability hinder their widespread adoption. Research is needed to develop robust electrode materials, high performance electrolytes, and advanced interphase engineering to overcome these limitations.

ML-driven BMS and DT frameworks have demonstrated significant potential for predictive maintenance, state of charge estimation, and real time performance optimization. However, most current models are limited by dataset availability, scalability, and generalization across battery chemistries and operational conditions. Future work should focus on integrating heterogeneous data sources, multi physics modeling, and real time adaptive algorithms to enable truly intelligent and resilient BMS.

Thermal runaway, degradation under high rate cycling, and mechanical stress induced failures remain critical safety concerns. While advances in thermal management, solid state electrolytes, and novel cooling strategies have mitigated some risks, comprehensive predictive models for multi cell modules and pack level integration are still lacking. Future research should address these gaps by combining materials innovation with AI-driven predictive modeling.

Transitioning from lab scale demonstrations to industrial-scale production poses significant challenges. Issues such as reproducibility of electrode microstructures, solid electrolyte fabrication, and quality control of DT-enabled BMS need systematic investigation. Research on scalable synthesis routes, advanced manufacturing techniques, and in line monitoring systems is essential for reliable mass production.

Environmental impact, recycling, and resource availability are critical for next generation battery adoption. Current studies often overlook end-of-life management, recycling efficiency, and LCA. Future directions should integrate circular economy principles, explore sustainable materials, and develop AI-driven lifecycle optimization strategies to ensure eco-friendly and economically viable battery solutions.

RL and safe-RL for charging/discharge scheduling, hybrid storage control, and thermal-aware policies remain promising yet underexplored avenues, particulary when combined with model-based constraints or digital twin in the loop evaluation.

Based on the identified gaps, several promising research directions can be envisioned:
\begin{itemize}
    \item Development of hybrid and multi-ion chemistries to simultaneously enhance energy density, power capability, and safety.
    \item Integration of advanced DL models with DT platforms for real time optimization, predictive maintenance, and autonomous BMS operation.
    \item Multi scale modeling frameworks that couple materials, cell, and pack level simulations for improved design and performance prediction.
    \item Implementation of solid-state electrolytes and novel electrode architectures to overcome dendrite formation, increase cycle life, and enhance safety.
    \item Adoption of sustainable and recyclable materials with AI-assisted lifecycle management for green and cost effective EV batteries.
\end{itemize}

Addressing these challenges will be pivotal in transitioning from conventional lithium-ion technology toward intelligent, high performance, and sustainable battery systems, thereby supporting the global electrification of mobility.

\section{\textbf{Conclusion}}
This survey has reviewed the state of the art in electrochemical energy storage and the emerging field of intelligent battery technologies for electric vehicles. While lithium-ion batteries remain the predominant choice, alternative chemistries, solid-state electrolytes, and novel electrode architectures offer promising pathways to overcome existing limitations in energy density, safety, and cycle life. Machine learning driven battery management systems, digital twin and large language models frameworks are proving instrumental in optimizing performance, predicting degradation, and extending operational longevity. Despite significant progress, several challenges remain, including scalability of advanced materials, lifecycle sustainability, integration of AI-based control systems, and real time predictive modeling under diverse operating conditions. Addressing these challenges through interdisciplinary research will be critical for the development of next generation, high performance, and eco friendly battery systems, ultimately supporting the widespread adoption of electric vehicles and the transition toward sustainable mobility.

\section*{Competing Interests}
The authors declare no conflict of interest.

\section*{Acknowledgements}
The authors acknowledge the Algerian Ministry of Higher Education and Scientific Research (MESRS).

\bibliographystyle{IEEEtran}

\end{document}